\documentclass[aps,floatfix,nofootinbib,showpacs,twocolumn]{revtex4} 
 
\usepackage{amsmath} 
\usepackage{graphicx}

\begin{document}

\title{Chemical potential and the gap equation}

\author{Huan Chen}
\affiliation{Department of Physics, Peking University, Beijing
100871, China}

\author{Wei Yuan}
\affiliation{Department of Physics, Peking University, Beijing
100871, China}

\author{Lei Chang}
\affiliation{Department of Physics, Peking University, Beijing
100871, China}

\author{Yu-Xin Liu\footnotemark[1]}
\affiliation{Department of Physics, Peking University, Beijing
100871, China}
\affiliation{Department of Physics and the State Key Laboratory of
Nuclear Physics and Technology, Peking University, Beijing 100871,
China} 
\affiliation{Center of Theoretical Nuclear Physics, National
Laboratory of Heavy Ion Accelerator, Lanzhou 730000, China}

\author{Thomas Kl\"ahn}

\affiliation{Physics Division, Argonne National Laboratory, Argonne,
Illinois 60439-4843, USA}

\author{Craig D. Roberts\footnotemark[1]}

\affiliation{Physics Division, Argonne National Laboratory, Argonne,
Illinois 60439-4843, USA}

\date{\today}

\begin{abstract}
In general the kernel of QCD's gap equation possesses a domain of analyticity upon which the equation's solution at nonzero chemical potential is simply obtained from the in-vacuum result through analytic continuation.  On this domain the single-quark number- and scalar-density distribution functions are $\mu$-independent.
This is illustrated via two models for the gap equation's kernel.  The models are alike in concentrating support in the infrared.  They differ in the form of the vertex but qualitatively the results are largely insensitive to the \textit{Ansatz}.  
In vacuum both models realise chiral symmetry in the Nambu-Goldstone mode and in the chiral limit, with increasing chemical potential, exhibit a first-order chiral symmetry restoring transition at $\mu \approx M(0)$, where $M(p^2)$ is the dressed-quark mass function.  
There is evidence to suggest that any associated deconfinement transition is coincident and also of first-order.
\end{abstract}


\maketitle

\section{Introduction}
QCD possesses a $U_B(1)$ symmetry, which is associated with a conserved charge; namely, baryon number, $Q_B$.  The finite baryon density theory is defined through the inclusion of a chemical potential, $\mu$, that is conjugate to $Q_B$.  The baryon number density is given by
\begin{equation}
\rho(\mu) = \frac{\partial P(\mu)}{\partial \mu}\,,
\end{equation}
where $P(\mu)$ is the thermodynamic pressure.  Naturally, in the confined domain the only contribution to the pressure from an excess of quarks over antiquarks is that generated by the presence of colour-singlet baryon bound-states.  That changes upon deconfinement.

\renewcommand{\thefootnote}{\fnsymbol{footnote}}

Confinement is a key emergent phenomenon in QCD.  Another is dynamical chiral symmetry breaking (DCSB), which is responsible, amongst many other things, for the large mass splitting between parity partners in the spectrum of light-quark hadrons, even though the relevant current-quark masses are small.  Neither of these phenomena is apparent in QCD's Lagrangian and yet they play a dominant role in determining the observable characteristics of real-world QCD.\footnotetext{\hspace*{-1em}$^\ast$\,Authors to whom relevant correspondence should be addressed: \mbox{liuyx@phy.pku.edu.cn} and/or cdroberts@anl.gov.}

\renewcommand{\thefootnote}{\arabic{footnote}}

In-medium; namely, for nonzero temperature and/or chemical potential, QCD will exhibit deconfinement and chiral symmetry restoration.  It is necessary to determine at which values of the intensive thermodynamical parameters these transitions take place, whether they are coincident, and to elucidate the response of the order parameters and also hadron properties to changes in the intensive parameters.  That information is important, for example, in constructing the equation of state and understanding the nature of matter in the core of dense astrophysical objects \cite{Klahn:2006ir}.  

One might look to numerical simulations of lattice-regularised QCD for this information. Such studies suggest that deconfinement and chiral symmetry restoration are coincident at $T\neq 0$ \protect\cite{Karsch:1998ua}.  However, a satisfactory algorithm for nonzero chemical potential is lacking \cite{Hands:2007by}.  Moreover, lattice-QCD has hitherto yielded little information on the in-medium behaviour of hadron properties \cite{Petreczky:2007ny}.

A natural starting point for an exploration of DCSB is QCD's gap equation.  There are numerous studies of this basic Dyson-Schwinger equation in-vacuum \cite{Maris:2003vk,Roberts:2007jh} and in-medium \cite{Roberts:2000aa}.  It is noteworthy that extant self-consistent studies of concrete models of QCD, which exhibit both confinement and DCSB in vacuum, possess coincident deconfinement and chiral symmetry restoration transitions; e.g., Refs.\,\cite{Bender:1996bm,Blaschke:1997bj,Bender:1997jf,Maris:2001rq,BashirNew}.  This result appears to follow from the crucial role played by the in-medium evolution of the dressed-quark self-energy in both transitions.  

Confinement herein is generally understood to be expressed through a violation of reflection positivity by coloured Schwinger functions \cite{Krein:1990sf}.  This is known to be a sufficient but not necessary condition.  

On the other hand, there is no ambiguity about DCSB.  It is signalled by the appearance of a gap equation solution in which the Dirac-scalar piece is nonzero in the chiral limit.  This effect owes primarily to a dense cloud of gluons that clothes a low-momentum quark \cite{Bhagwat:2007vx,Roberts:2007ji}.  DCSB is the single most important mass generating mechanism for light-quark hadrons; e.g., one can identify it as being responsible for roughly 98\% of a proton's mass.  A system in which the gap equation's DCSB solution is the favoured ground state is said to realise chiral symmetry in the Nambu-Goldstone mode.  The antithesis is termed a Wigner-Weyl realisation of the symmetry.  

Our theme is $\mu\neq 0$ because, compared with nonzero temperature, this is the poorer understood domain.  Models and truncations of QCD can therefore provide important information.  Simple models suggest that a first-order chiral symmetry restoring phase transition occurs at $\mu=\mu_c^{\chi} \sim M_Q$, where $M_Q$ is a constituent-quark mass-scale.  This happens, e.g., in Refs.\,\cite{Blaschke:1997bj,Bender:1997jf}, which exhibit a coincident deconfinement transition.  It is noteworthy that below $\mu_c^{\chi}$ the structure of the ground state can alter gradually but its qualitative character is unchanged.  

Our study complements existing work toward understanding $\mu\neq 0$ in QCD.  In addition to those noted above, kindred studies of: chiral symmetry restoration are described in Refs.\,\cite{Harada:1998zq,Liu:2003bf,Zong:2005mm,zongPLB}; the response of hadron properties, in Refs.\,\cite{Maris:1997eg,Liu:2001em,Liu:2003jx,Chang:2005ay}; the possible realisation of colour superconductivity, in Refs.\,\cite{Bloch:1999vk,Nickel:2006vf,Nickel:2006kc,Yuan:2006yf}; and quark stars, in Ref.\,\cite{Blaschke:1998hy,Jaikumar:2005hy}.  One novelty of our study is a  consideration of the effect of dressing the quark-gluon vertex.

This article is organised as follows.  In Sec.\,\ref{sec:formal} we describe the $\mu\neq 0$ gap equation and its solution.  This enables us to explain the notion of a domain of analyticity in $\mu$, upon which the in-medium Schwinger functions can be obtained from their vacuum values via straightforward analytic continuation, and the number- and scalar-densities, e.g., are $\mu$-independent.  In Sec.\,\ref{sec:models} we introduce our models for the gap equation's kernel.  The primary model can be described as lying within the class of rainbow truncations of the gap equation.  It can be systematically improved.  The second differs in that we employ a vertex \emph{Ansatz}, which is interpreted as modelling effects of dressing.  Section~\ref{sec:numerical} is extensive.  It reports numerical results obtained in the chiral limit and includes a brief discussion of confinement expressed through the violation of reflection positivity \cite{Krein:1990sf,gj81,Osterwalder:1973dx,Osterwalder:1974tc}.  We recapitulate and conclude in Sec.\,\ref{sec:summary}.

\section{Formal structure and analysis}
\label{sec:formal}
Herein we elucidate the influence of $\mu$ on the nature of the solutions to QCD's gap equation, which reads\footnote{In our Euclidean metric:  $\{\gamma_\rho,\gamma_\sigma\} = 2\delta_{\rho\sigma}$; $\gamma_\rho^\dagger = \gamma_\rho$; $\gamma_5= \gamma_4\gamma_1\gamma_2\gamma_3$; $a \cdot b = \sum_{i=1}^4 a_i b_i$, $\vec{a}\cdot \vec{b}=\sum_{i=1}^3 a_i b_i$; and $P_\rho$ timelike $\Rightarrow$ $P^2<0$.}
\begin{equation}
 S(p;\mu)^{-1}= Z_2 (i\vec{\gamma}\cdot \vec{p} + i \gamma_4 (p_4+i\mu) + m^{\rm bm}) + \Sigma(p;\mu)\,, \label{gendse}
\end{equation}
with the renormalised self energy expressed as
\begin{eqnarray}
\nonumber 
\Sigma(p;\mu ) &=& Z_1 \int^\Lambda_q\! g^2(\mu) D_{\rho\sigma}(p-q;\mu) \\
& & \times \frac{\lambda^a}{2}\gamma_\rho S(q;\mu) \Gamma^a_\sigma(q,p;\mu) , \label{gensigma}
\end{eqnarray}
where $\int^\Lambda_q$ represents a translationally invariant regularisation of the integral, with $\Lambda$ the regularisation mass-scale, $D_{\rho\sigma}(k;\mu)$ is the dressed-gluon propagator, $\Gamma^a_\sigma(q,p;\mu)$ is the dressed-quark-gluon vertex, and $m^{\rm bm}$ is the $\Lambda$-dependent current-quark bare mass.  The quark-gluon-vertex and quark wave function renormalisation constants, $Z_{1,2}(\zeta^2,\Lambda^2)$, depend on the renormalisation point, $\zeta$, the regularisation mass-scale and the gauge parameter.  A nonzero chemical potential introduces no divergences in addition to those present in the $\mu=0$ theory.  Hence, the renormalisation constants determined at $\mu=0$ are completely sufficient in-medium.

At $\mu=0$ the gap equation's solution has the form 
\begin{eqnarray} 
 \bar S(p)^{-1} & = & i \gamma\cdot p \, \bar A(p^2,\zeta^2) + \bar B(p^2,\zeta^2) \,.
%
\label{sinvp0} 
\end{eqnarray}
and the mass function $\bar M(p^2)=\bar B(p^2,\zeta^2)/\bar A(p^2,\zeta^2)$ is renormalisation point independent.  (Hereafter, when emphasis is useful or necessary, an overscore denotes a quantity calculated in-vacuum; viz., with $\mu = 0$.)  The propagator is obtained from Eq.\,(\ref{gendse}) augmented by the condition
\begin{equation}
\label{renormS} \left.\bar S(p)^{-1}\right|_{p^2=\zeta^2} = i\gamma\cdot p + m(\zeta)\,,
\end{equation}
where 
\begin{equation}
m(\zeta) = [Z_2(\zeta^2,\Lambda^2)/Z_4(\zeta^2,\Lambda^2)]  \, m^{\rm bm}(\Lambda),
\end{equation}
is the renormalised mass, with $Z_4$ the Lagrangian mass renormalisation constant.  The chiral limit means $\hat m = 0$, where $\hat m$ is the renormalisation-group-invariant light-quark current-mass \cite{Maris:1997hd}.

Poincar\'e covariance is lost at $\mu \neq 0$. In this case the gap equation's solution can assume the general form \cite{Rusnak:1995ex}
\begin{eqnarray} 
\nonumber 
\lefteqn{ 
S(p;\mu)^{-1} = i \vec{\gamma}\cdot \vec{p} \, A(p^2,p\cdot u,\zeta^2) }\\
%
&+&  i \gamma_4(p_4+i\mu) \, C(p^2,p\cdot u,\zeta^2) + B(p^2,p\cdot u,\zeta^2)  \,,
%
\label{sinvp} 
\end{eqnarray}
where we have written $u=(\vec{0},i\mu)$.  Rotational invariance remains.  The propagator itself can be written
\begin{eqnarray} 
\nonumber 
S(p;\mu) &= &i \vec{\gamma}\cdot \vec{p} \, \sigma_A(p^2,p\cdot u,\zeta^2) + \sigma_B(p^2,p\cdot u,\zeta^2)  \\
&+& i \gamma_4(p_4+i\mu) \,\sigma_C(p^2,p\cdot u,\zeta^2)\,.\label{sitselfp} 
%
\end{eqnarray}

Consider now the kernel of the gap equation; viz., the integrand in Eq.\,(\ref{gensigma}), and chemical potential in the neighbourhood of $\mu=0$.  So long as the kernel is analytic on a set of nonzero measure in the complex-$q_4^2$ plane that includes the nonnegative real-$q_4^2$ axis, then the dressed-quark propagator is obtained from that in-vacuum through a straightforward analytic continuation.  As recognised in Ref.\,\cite{Zong:2005mm}, this procedure remains valid so long as no singularity in the kernel is contained or moves within the contour 
\begin{equation}
\begin{array}{lcl}
\gamma & = & \lim_{R\to \infty} \cup_{i=1}^4\gamma_i\,,\\
\gamma_1 &=& \{ z(t)= t, -R\leq t \leq R\}\,,\\
\gamma_2 &=& \{ z(t)= R+it, 0\leq t \leq \mu \}\,,\\
\gamma_3 &=& \{ z(t)= t+i\mu, R\geq t \geq -R\}\,,\\
\gamma_4 &=& \{ z(t)= -R+it, \mu \geq  t \geq 0 \}\,.
\end{array}
\end{equation}

If one evaluates the integral in Eq.\,(\ref{gensigma}) using a free gauge boson propagator, a bare vertex and a free fermion propagator with mass $M$, then the conditions above are satisfied for $\mu<M$.  Plainly, this is a statement that nothing occurs in perturbation theory unless the chemical potential exceeds the Fermi energy. 

On the other hand, suppose that nonperturbatively $\exists\mu_a$ such that for $\mu < \mu_a$ the analyticity condition is satisfied, then 
\begin{equation}
\label{SmuS0}
\forall \mu < \mu_a :\; S(p;\mu) = \bar S(p+u)\,,
\end{equation}
from which it follows that 
\begin{equation}
\forall \mu < \mu_a :\;
\left\{\begin{array}{lcl}
A(p^2,p\cdot u) & \equiv & C(p^2,p\cdot u)\\
 &  = & \bar A((p+u)^2)\,,\\
B(p^2,p\cdot u) & = & \bar B((p+u)^2)\,.
\end{array} \right.
\end{equation}
DSE models have been explored in which $\mu_a=0$ \cite{Blaschke:1997bj,Bender:1997jf} and $\mu_a \neq 0$ but small \cite{zongPLB}.  Variants of the Nambu--Jona-Lasinio model can be constructed in which $\mu_a > M_Q$, where $M_Q$ is the model's constituent-quark mass-scale \cite{Buballa:2003qv}. 

The gap equation is one of a system of Dyson-Schwinger equations (DSEs) \cite{Maris:2003vk,Roberts:2007ji,Roberts:2000aa}, from whose structure one may infer that $\exists\, \mu_a^{\cal I}\geq 0$, an infimum, such that $\forall \mu < \mu_a^{\cal I}$ every Schwinger function can simultaneously be obtained via analytic continuation of its in-vacuum form.  On this domain, in addition to Eq.\,(\ref{SmuS0}) one has, e.g., 
\begin{eqnarray}
D_{\rho\sigma}(k;\mu) &=& \bar D_{\rho\sigma}(k)\,, \label{Dkmu}\\
\Gamma^a_\rho(p,q;\mu) & = & \bar \Gamma^a_\rho(p+u,q+u)\,. \label{vertexmu}
\end{eqnarray}

Another two important values of the chemical potential are those above which quarks are no longer confined and chiral symmetry is no longer realised in the Nambu-Goldstone mode.  We denote them, respectively, by $\mu_c^d$ and $\mu_c^\chi$.  There are models in which $\mu_c^d = \mu_c^\chi$ \cite{Bender:1997jf,Blaschke:1997bj}.  That might also be the case in QCD.  We write 
\begin{equation}
\mu_c = \inf(\mu_c^d,\mu_c^\chi)\,.
\end{equation}
It is noteworthy that the analytic properties of the dressed-quark propagator, for example, change abruptly at $\mu_c$. 

It is relevant to list a few additional results that obtain on the domain $0 <\mu < \mu_c^{\cal I}$, 
\begin{equation}
\mu_c^{\cal I}=\inf(\mu_a^{\cal I},\mu_c)\,.
\end{equation}
We define the quark number density 
\begin{eqnarray}
\label{nqmu}
n_q(\mu) & = & 2 N_c N_f Z_2 \int\frac{d^3 p}{(2\pi)^3} \, f_1(|\vec{p}|;\mu)\,, \\
\label{nqmuf1}
f_1(|\vec{p}|;\mu) & = & \frac{1}{4\pi} \int_{-\infty}^\infty \! dp_4 \, {\rm tr}_{\rm D} (-\gamma_4) S(p;\mu)\,.
\end{eqnarray}
For $\mu <\mu_c^{\cal I}$ one can be confident that Eq.\,(\ref{SmuS0}) is satisfied and hence
\begin{equation}
\forall \mu < \mu_c^{\cal I}:\; f_1(|\vec{p}|;\mu) = 0 \; \Rightarrow \; n_q(\mu) =0 \,;
\end{equation}
namely, the system exhibits no excess of quarks over antiquarks.\footnote{For a noninteracting single-species Fermi gas this situation pertains for $\mu<M$, where $M$ is the fermion mass.}  Plainly, on this same domain there is no contribution to the thermodynamic pressure from an excess of quarks over antiquarks.

The vacuum quark condensate can be expressed \cite{Langfeld:2003ye}
\begin{eqnarray}
\label{qbq0mu}
-\langle \bar q q \rangle_\zeta^0(\mu) & = & 2 N_c Z_4 \int\frac{d^3 p}{(2\pi)^3} \, f_2(|\vec{p}|;\mu)\,, \\
\label{qbq0muf2}
f_2(|\vec{p}|;\mu) & = & \frac{1}{4\pi} \int_{-\infty}^\infty\! dp_4 \, {\rm tr}_{\rm D} S(p;\mu)\,.
\end{eqnarray}
The reasoning used above leads to the conclusion that
\begin{equation}
\forall \mu < \mu_c^{\cal I}:\; f_2(|\vec{p}|;\mu) = \bar f_2(|\vec{p}|) \; \Rightarrow \; \langle \bar q q \rangle_\zeta^0(\mu) = \overline{\langle \bar q q \rangle_\zeta^0 }\,;
\end{equation}
i.e., on this domain the vacuum quark condensate is independent of $\mu$.

It will be evident at this point that the simplest situation to understand is that with $\mu_c \leq \mu_a^{\cal I}$.  As we have just explained, in this case the system's ground state is insensitive to increasing $\mu$ until the phase boundary is encountered.  We observed above, however, that instances exist in which $\mu_a < \mu_c$.  Such are possibly the most physically relevant cases, in which order parameters and observables may exhibit a qualitative change at $\mu=\mu_a$ and thereafter evolve continuously on the domain $\mu\in (\mu_a,\mu_c)$.  This is illustrated by Ref.\,\cite{Blaschke:1997bj}, in which $\mu_a=0$ and $n_q(\mu)\equiv 0$ on $\mu\in [0,\mu_c)$ but the condensate grows with increasing $\mu$.

\section{Gap equation models}
\label{sec:models}
Significant effort continues to be expended on determining the precise nature of the kernel of QCD's gap equation.  A dialogue between DSE studies and results from numerical simulations of lattice-regularised QCD is providing important information; e.g., Refs.\,\cite{Bhagwat:2003vw,Alkofer:2003jj,Bhagwat:2004hn,Bhagwat:2004kj,Bhagwat:2006tu,%
Kizilersu:2006et,Kamleh:2007ud,Boucaud:2008ky,Cucchieri:2008fc}.  This body of work can be used to formulate reasonable \textit{Ans\"atze} for the dressed-gluon propagator and dressed-quark-gluon vertex in Eq.\,(\ref{gensigma}).  That is how we proceed.  The approach can be particularly fruitful in connection with QCD at $\mu>0$ since little is rigorously known about that theory.

In connection with employment of these \emph{Ans\"atze} it is worth remarking that the DSEs admit at least one nonperturbative symmetry preserving truncation scheme \cite{Bhagwat:2004hn,Munczek:1994zz,Bender:1996bb}, which has enabled the proof of numerous exact results \cite{Maris:1997hd,Bicudo:2003fp,Holl:2004fr,Holl:2005vu,McNeile:2006qy,Bhagwat:2007ha,%
Ivanov:1997yg,Ivanov:1998ms,Bhagwat:2006xi}.  It also provides a starting point for the formulation of reliable models that can be used to illustrate those results and make predictions with readily quantifiable errors \cite{Roberts:2007jh,Maris:1997tm,Maris:1999nt,Bloch:2002eq,Maris:2002mt,Eichmann:2008ae}.  

The \emph{Ans\"atze} are typically implemented by writing
\begin{eqnarray}
\nonumber \lefteqn{Z_1 g^2 D_{\rho \sigma}(p-q) \Gamma_\sigma^a(q,p)} \\
& =& {\cal G}((p-q)^2) \, D_{\rho\sigma}^{\rm free}(p-q) \frac{\lambda^a}{2}\Gamma_\sigma(q,p)\,, \label{KernelAnsatz}
\end{eqnarray}
wherein $D_{\rho \sigma}^{\rm free}(\ell)$ is the Landau-gauge free gauge boson propagator, ${\cal G}(\ell^2)$ is a model effective-interaction and $\Gamma_\sigma(q,p)$ is a vertex \textit{Ansatz}.  In one widely used approach to in-vacuum physics, ${\cal G}(\ell^2)$ is chosen such that the one-loop renormalisation group behaviour of QCD is preserved and the vertex is written
\begin{equation}
\label{rainbowV}
\Gamma_\sigma(q,p) = \gamma_\sigma\,.
\end{equation}
This is the basis for a renormalisation-group-improved rainbow-ladder truncation of QCD's DSEs  \cite{Roberts:2007jh,Maris:1997tm,Maris:1999nt,Bloch:2002eq,Maris:2002mt,Eichmann:2008ae}.  

If one is interested solely in the gap equation and not concerned with a symmetry preserving treatment of bound-state properties, then \emph{Ans\"atze} for the vertex can be employed whose diagrammatic content is unknown.  The class of such models that has hitherto been employed can be characterised by \cite{Ball:1980ay}
\begin{eqnarray}
\label{bcvtx}
\nonumber \lefteqn{i\Gamma_\sigma(k,\ell)  =
i\Sigma_A(k^2,\ell^2)\,\gamma_\sigma + (k+\ell)_\sigma }\\
&\times &
\left[\frac{i}{2}\gamma\cdot (k+\ell) \,
\Delta_A(k^2,\ell^2) + \Delta_B(k^2,\ell^2)\right] \!,
\end{eqnarray}
where 
\begin{eqnarray}
\Sigma_F(k^2,\ell^2)& = &\frac{1}{2}\,[F(k^2)+F(\ell^2)]\,,\;\\
\Delta_F(k^2,\ell^2) &=&
\frac{F(k^2)-F(\ell^2)}{k^2-\ell^2}\,,
\label{DeltaF}
\end{eqnarray}
with $F= A, B$; viz., the scalar functions in Eq.\,(\ref{sinvp0}).  This \emph{Ansatz} satisfies the vector Ward-Takahashi identity and on $0 < \mu <\mu_c^{\cal I}$ its in-medium form is obtained through Eq.\,(\ref{vertexmu}).  For $\mu \geq \mu_c^{\cal I}$, 
Eq.\,(\ref{bcvtx}) generalises to \cite{Maris:2000ig}
\begin{eqnarray}
\nonumber
\lefteqn{i\Gamma_\sigma(k,\ell;\mu) = i \Sigma_A(k,\ell;\mu) \gamma^\perp_\sigma + i \Sigma_C(k,\ell;\mu) \gamma^\|_\sigma}\\
\nonumber
&+&(\tilde k+\tilde\ell)_\sigma \left[ 
\frac{i}{2}\gamma^\perp\cdot (\tilde k+\tilde\ell) \Delta_A(\tilde k,\tilde\ell;\mu) \right.\\
&+& \left.
\frac{i}{2}\gamma^\|\cdot (\tilde k+\tilde\ell) \Delta_C(\tilde k,\tilde\ell;\mu)
+ \Delta_B(\tilde k,\tilde\ell;\mu)
\right], \label{bcvtxmu}
\end{eqnarray} 
where $\tilde k = k+u$, 
\begin{eqnarray}
\Sigma_F(k,\ell;\mu) & = & \frac{1}{2} \left[ F(\vec k^2,k_4;\mu)+F(\vec \ell^2,\ell_4;\mu)\right],\\
\Delta_F(\tilde k,\tilde \ell;\mu) & = &  \frac{F(\vec k^2,k_4;\mu)-F(\vec \ell^2,\ell_4;\mu)}{\tilde k^2-\tilde \ell^2},\\
\end{eqnarray}
and we have defined $\gamma^\perp = \gamma - \hat u \gamma\cdot \hat u $, $\gamma^\| = \hat u \gamma\cdot \hat u$, with $\hat u^2=1$.  

One has a Slavnov-Taylor identity for the quark-gluon vertex in QCD, not a Ward-Takahashi identity.  Hence, Eqs.\,(\ref{bcvtx}) and (\ref{bcvtxmu}) are not necessarily an improvement over Eq.\,(\ref{rainbowV}).  A comparison between results obtained with the different \emph{Ans\"atze} is nevertheless useful in identifying those which might be robust.  It is noteworthy that our vertex \emph{Ans\"atze} do not exhibit kinematic singularities \cite{Burden:1993gy}.

Herein we employ a simplified form of the renormalisation-group-improved effective interaction in Refs.\,\cite{Roberts:2007jh,Maris:1997tm,Maris:1999nt,Bloch:2002eq,Maris:2002mt,%
Eichmann:2008ae}; viz., we retain only that piece which expresses the long-range behaviour ($s=k^2$):
\begin{equation}
\label{IRGs}
\frac{{\cal G}(s)}{s} = \frac{4\pi^2}{\omega^6} \, D\, s\, {\rm e}^{-s/\omega^2}.
\end{equation}
This is a finite width representation of the form introduced in Ref.\,\cite{mn83}, which has been rendered as an integrable regularisation of $1/k^4$ \cite{mm97}.  Equation~(\ref{IRGs}) delivers an ultraviolet finite model gap equation.  Hence, the regularisation mass-scale can be removed to infinity and the renormalisation constants set equal to one.  We specify the $\mu\neq 0$ \emph{Ansatz} via Eq.\,(\ref{Dkmu}).  In making this modest and practical simplification we nevertheless proceed beyond the $\mu\neq 0$ studies of, e.g., Refs.\,\cite{Blaschke:1997bj,Bender:1997jf,%
Maris:1997eg,Liu:2001em,Liu:2003jx,Chang:2005ay,Bloch:1999vk}.  Studies are underway with the complete interaction of Refs.\,\cite{Roberts:2007jh,Maris:1997tm,Maris:1999nt,Bloch:2002eq,Maris:2002mt,%
Eichmann:2008ae}.  

\section{Numerical Results}
\label{sec:numerical}
The active parameters in Eq.\,(\ref{IRGs}) are $D$ and $\omega$.  However, they are not independent.  In reconsidering a renormalisation-group-improved rainbow-ladder fit to a selection of ground state observables \cite{Maris:1999nt}, Ref.\,\cite{Maris:2002mt} noted that a change in $D$ is compensated by an alteration of $\omega$.  This feature has further been elucidated and exploited in Ref.\,\cite{Eichmann:2008ae}.  With the interaction specified by Eqs.\,(\ref{KernelAnsatz}), (\ref{rainbowV}) and (\ref{IRGs}), fitted in-vacuum low-energy observables are approximately constant along the trajectory\footnote{The value of $m_g$ is typical of the mass-scale associated with nonperturbative gluon dynamics.}
\begin{equation}
\label{gluonmass}
\omega D  = (0.8 \, {\rm GeV})^3 =: m_g^3\,.
\end{equation}
Herein, we employ $\omega=0.5\,$GeV.

\begin{table}[t]
\caption{Results obtained for selected quantities with the vertex and $D$ parameter value indicated:
$A(0)$, $M(0)$ are $p=0$ in-vacuum values of the scalar functions defined in connection with Eq.\,(\ref{sinvp0}); the vacuum quark condensate is defined in Eq.\,(\ref{qbq0mu}); and $M$, $\Gamma$ are discussed in connection with Eq.\,(\protect\ref{stinglSN}).
All calculated quantities in GeV except $D$, in ${\rm GeV}^2$, and $A(0)$, dimensionless.
Unless otherwise noted, all calculations reported herein were performed in the chiral limit.
\label{Table:Para1} 
}
\begin{center}
\begin{tabular*}
{\hsize}
{|l@{\extracolsep{0ptplus1fil}}
|c@{\extracolsep{0ptplus1fil}}
|c@{\extracolsep{0ptplus1fil}}
|c@{\extracolsep{0ptplus1fil}}
|c@{\extracolsep{0ptplus1fil}}
|c@{\extracolsep{0ptplus1fil}}
|c@{\extracolsep{0ptplus1fil}}
|c@{\extracolsep{0ptplus1fil}}
|c@{\extracolsep{0ptplus1fil}}|} \hline
Vertex & $D$ & $A(0)$ & $M(0)$ & $-(\langle \bar q q \rangle^0)^{1/3}$ & $M_1$ & $\Gamma_1$  & $M_2$ & $\Gamma_2$  \\\hline
Eq.\,(\ref{rainbowV}) & 1 & 1.3 & 0.40 & 0.25 & 0.53 & 0.17 & 0.82 & 0.0~~ \\\hline
Eq.\,(\ref{bcvtx}) & $\frac{1}{2}$ & 1.1 & 0.28 & 0.26 & 0.30 & 0.0\,~ & 0.80 & 0.20 \\\hline
\end{tabular*}
\end{center}
\end{table}

It is not possible to explore whether the behaviour expressed in connection with Eq.\,(\ref{gluonmass}) is also realised with the interaction specified by Eqs.\,(\ref{KernelAnsatz}), (\ref{bcvtx}) and (\ref{IRGs}) because Eq.\,(\ref{bcvtx}) cannot yet be used in the implementation of a symmetry preserving DSE truncation.  Hence, we keep $\omega=0.5\,$GeV and choose $D=0.5\,$GeV$^2$, a combination that produces the chiral-limit vacuum quark condensate in Row~3 of Table~\ref{Table:Para1}.

\subsection{Rainbow truncation}
It is straightforward to solve the gap equation with the interaction specified by Eqs.\,(\ref{KernelAnsatz}), (\ref{rainbowV}) and (\ref{IRGs}), the regularisation mass-scale removed to infinity and the renormalisation constants set equal to one.  In Table~\ref{Table:Para1} we report in-vacuum values characteristic of the solution.  

Our first novel goal was to determine $\mu_a$ in Eq.\,(\ref{SmuS0}).  In a system that supports dynamical chiral symmetry breaking that can be done by analysing the following configuration-space Schwinger function \cite{Hawes:1993ef}: with $\tau>0$, 
\begin{eqnarray}
\label{tauSchwinger}
\Delta(\tau,\mu) &= & \int \frac{d^4p}{(2\pi)^4}\,{\rm e}^{i \vec{p}\cdot \vec{x} + i p_4 \tau} \, \delta(\vec{p}) \, \sigma_B(p;\mu)\,.
\end{eqnarray}
Our results for a range of values of $\mu$ are depicted in Fig.\,\ref{fig:SF1}.  NB.\  In the absence of DCSB; i.e., when chiral symmetry is realised in the Wigner mode, $\sigma_B \equiv 0$ in the chiral limit.  In the Wigner mode
\begin{equation}
\mu_a^W = 0
\end{equation}
because there is no dynamically generated mass-scale that can raise the Fermi energy of the massless theory.

\subsubsection{Reflection positivity}
\label{Reflection}
\begin{figure}[t]
\centerline{
\includegraphics[scale=0.810] {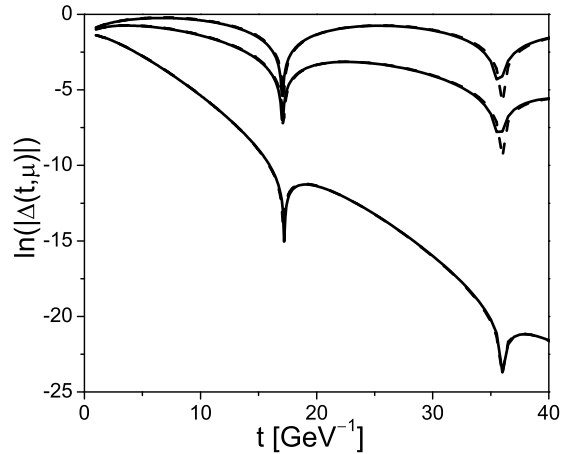}} 
\centerline{
\includegraphics[scale=0.810] {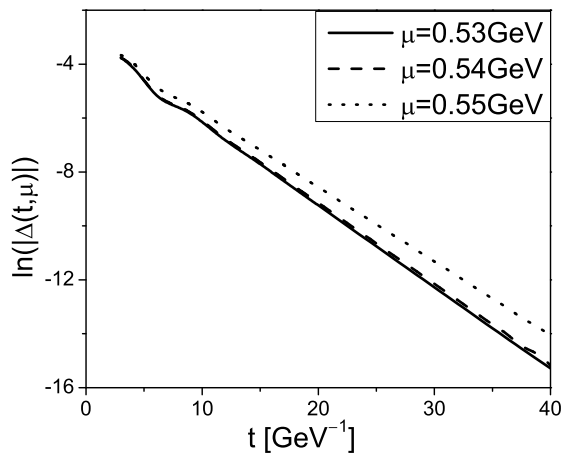}} 
\caption{Schwinger function in Eq.\,(\protect\ref{tauSchwinger}) computed from the gap equation defined via Eqs.\,(\ref{KernelAnsatz}), (\ref{rainbowV}) and (\ref{IRGs}).  \emph{Upper panel} -- from bottom to top, $\mu = 0$, $0.4\,$GeV, $0.5\,$GeV.  Solid curves: our numerical results.  Dashed curves: fit obtained via Eq.\,(\protect\ref{Deltacc}), with $M$, $\Gamma$ given in Table~\protect\ref{Table:Para1}.  \emph{Lower panel} -- $\mu_a < \mu = 0.53$, $0.54$, $0.55\,$GeV. \label{fig:SF1}}
\end{figure}

In order to explain Fig.\,\ref{fig:SF1} we must first digress.  Recall that in a quantum field theory defined by a Euclidean measure \cite{gj81}, the Osterwalder-Schrader axioms \cite{Osterwalder:1973dx,Osterwalder:1974tc} are five conditions which any moment of this measure ($n$-point Schwinger function) must satisfy if it is to have an analytic continuation to Minkowski space and hence an association with observable quantities. One of these is ``OS3,'' the axiom of reflection positivity, which is violated if the Schwinger function's Fourier transform to configuration space is not positive definite. The space of observable asymptotic states is spanned by eigenvectors of the theory's infrared Hamiltonian and no Schwinger function that breaches OS3 has a correspondent in this space.  Consequently, the violation of OS3 is a sufficient condition for confinement.  This connection has long been of interest and is explained, e.g., in Sec.\,2 of Ref.\,\cite{Roberts:2007ji}.  

For a free fermion of mass $M$
\begin{equation}
\label{SfreeM}
S^{\rm free}(p) = \frac{1}{i\gamma\cdot p + M}
\end{equation}
and one finds 
\begin{equation}
\label{Deltafree}
\Delta^{\rm free}(\tau,\mu) = \frac{1}{2}\, {\rm e}^{-\tau (M-\mu)} \theta(M-\mu)\,,
\end{equation}
which is positive definite for $\mu<M$.  Hence, OS3 is satisfied on this domain and, indeed, there is certainly a well-defined single-particle asymptotic state associated with this Schwinger function.  At $\mu = M$, on the other hand, this single-particle Schwinger function vanishes.  That happens because the Dirac mass gap has vanished and the Fermi energy is breached.  At this point an instability appears in the theory, one that is associated with real-particle production.  Plainly, as we indicated above, $\mu_a=M$ for a free fermion.

Suppose instead that one encounters a theory in which the dressed-quark propagator assumes the form
\cite{Habel:1990tw,Stingl:1994nk,Bhagwat:2002tx}
\begin{equation}
\label{stinglS}
S_{cc}(p)= \frac{1}{2}\left[ \frac{1}{i \gamma\cdot p + z} + 
\frac{1}{i \gamma\cdot p + z^\ast} \right],\;z=M + i \Gamma\,,
%
\end{equation}
a function with conjugate poles in the complex-$p_4$ plane; viz., poles at $\pm (i M \pm \Gamma)$ when $\vec{p}=0$.  In the absence of a chemical potential, this yields
\begin{equation}
\Delta_{cc}(\tau) = \frac{1}{2} \, {\rm e}^{-M \tau}\, \cos(\Gamma\tau)\,,
\end{equation}
a function that possesses infinitely many, regularly spaced zeros.  Hence, OS3 is violated and thus the fermion described by this Schwinger function has
no correspondent in the space of observable asymptotic states.  

When $\mu\neq 0$ one has
\begin{equation}
\label{Deltacc}
\Delta_{cc}(\tau,\mu) = \frac{1}{2} \, {\rm e}^{-(M-\mu) \tau}\, \cos(\Gamma\tau )\, \theta(M-\mu)\,.
\end{equation}
This is a single-particle Schwinger function with a mass gap that vanishes when the chemical potential exceeds $\Re(z= M + i \Gamma) = M$, which is the magnitude of the imaginary part of the pole's position in the complex-$p_4$ plane.

It is conceivable that a theory might produce
\begin{equation}
\label{stinglSN}
S_{N}(p)= \sum_{i=1}^N \,\frac{r_i}{2}\left[ \frac{1}{ i\gamma\cdot p +  z_i} + 
\frac{1}{i \gamma\cdot p + z_i^\ast} \right]\,,
\end{equation}
where, without loss of generality, $0\leq\Re(z_i)<\Re(z_{i+1})$, $i=1,2,\ldots,N-1$; namely, a single-fermion Schwinger function possessing a sequence of $N$ real and/or complex-conjugate poles with associated residues $r_i$.  If this theory preserves the ultraviolet behaviour of QCD, Eq.\,(\ref{renormS}), then
\begin{equation}
\sum_{i=1}^N \,r_i(\zeta) = 1 \,,\; \sum_{i=1}^N \,r_i(\zeta) \,\Re(z_i) = m(\zeta)\,.
\end{equation}
In this case
\begin{equation}
\label{DeltaN}
\Delta_{N}(\tau,\mu) = \sum_{i=1}^N \frac{r_i}{2} \, {\rm e}^{-(\Re(z_i)-\mu) \tau}\, \cos(\Im(z_i)\tau )\, \theta(\Re(z_i)-\mu)\,,
\end{equation}
which at fixed $\tau\geq 0$ is a piecewise continuous function of $\mu$.  Plainly, one can therefore identify 
\begin{equation}
\label{muageneral}
\mu_a = \Re(z_1)\,;
\end{equation}
i.e., with the location of the first discontinuity.  Equation~(\ref{stinglSN}) characterises a class of single-fermion Schwinger functions that contains all the cases we've considered.  Hence, Eq.\,(\ref{muageneral}) is a general statement valid for all theories in this class.

\subsubsection{Domain of analyticity}
We now return to Fig.\,\ref{fig:SF1}.  The curves coincide with a Schwinger function of the types discussed in Eqs.\,(\ref{stinglS}) and (\ref{stinglSN}).  The evolution with $\tau$ is dominated by the pole for which $\Re(z)$ is the smallest, and our fit yields $z_1 = M_1+i\Gamma_1$, with the values given in Table~\ref{Table:Para1}.  The quality of the fit is good.  In this case, 
\begin{equation}
\mu_a^{\rm rb} = 0.53\,\,{\rm GeV}.
\end{equation}
The lower panel depicts results obtained with $\mu \gtrsim \mu_a^{\rm rb}$.  They are dramatically different from those in the upper panel.  It will be plain from Sec.\,\ref{Reflection} that the evolution with $\tau$ indicates the presence of another singularity, $z_2=M_2+ i \Gamma_2$ in Eq.\,(\ref{stinglSN}).  The fitted values are given in Table~\ref{Table:Para1}.  No imaginary part appears in this instance.

\begin{figure}[t]
\centerline{
\includegraphics[scale=0.810] {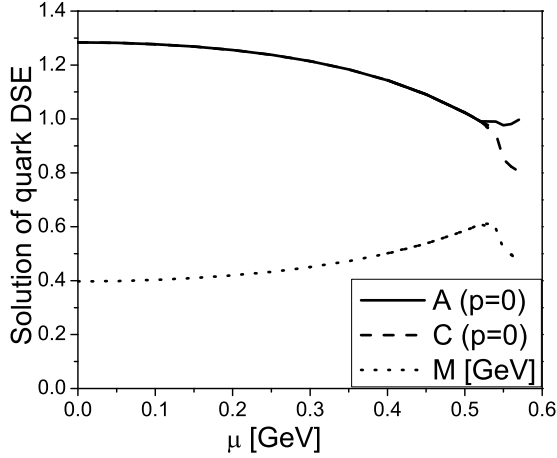}} 
\caption{Evolution with $\mu$ of the dimensionless quantities $A(p=0;\mu)$, $C(p=0;\mu)$, and $M(p=0;\mu)$ measured in GeV, all computed from the gap equation defined via Eqs.\,(\ref{KernelAnsatz}), (\ref{rainbowV}) and (\ref{IRGs}). \label{fig:ABCM}}
\end{figure}

In Fig.\,\ref{fig:ABCM} we present the evolution with $\mu$ of the scalar functions in the dressed-quark propagator evaluated at $p=0$.  Consistent with the preceding discussion; viz., Eq.\,(\ref{SmuS0}), $A(p=0;\mu)=C(p=0;\mu)$ for $\mu<\mu_a^{\rm rb}$.  Nonetheless, all the plotted quantities evolve with $\mu$.  In particular, $M(0;\mu)$ is monotonically increasing on this domain, as it is in all models that preserve the momentum-dependence of dressed-quark self-energies \cite{Roberts:2000aa}.

\begin{figure}[t]
\centerline{
\includegraphics[scale=0.810] {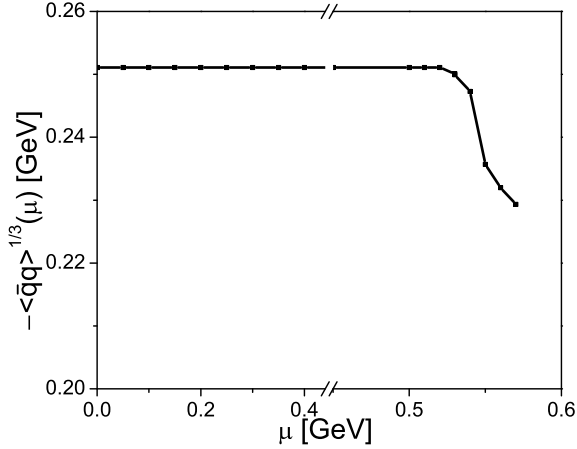}} 
\caption{Chemical potential dependence of the vacuum quark condensate computed from the solutions of the gap equation defined via Eqs.\,(\ref{KernelAnsatz}), (\ref{rainbowV}) and (\ref{IRGs}).  \label{fig:condensate}}
\end{figure}

Figure~\ref{fig:condensate} displays the behaviour of the chiral-limit vacuum quark condensate, Eq.\,(\ref{qbq0mu}).  Despite the fact that the scalar functions in $S(p;\mu)$ evolve with $\mu$, the condensate is constant on $\mu<\mu_a^{\rm rb}$.

\subsubsection{Phase transition}
\label{RainbowPT}
In any study relating to QCD at nonzero chemical potential it is crucial to determine the point at which the theory makes a transition between the Nambu-Goldstone and Wigner realisations of chiral symmetry.  All reasonable models support both these phases and those considered herein are not exceptions.  An exposition of the nature of these phases is presented in Ref.\,\cite{Chang:2006bm}.

In a consistent rainbow truncation the transition can be studied by considering the pressure owing to dressed-quarks obtained through the ``steepest descent'' approximation; namely, 
\begin{equation}
\label{pSigma}
p_{\Sigma}(\mu) =  {\rm TrLn}\left[S^{-1}\right] - \frac{1}{2}{\rm Tr}\left[\Sigma\,S\right],
\end{equation}
where $S$ is the solution of the gap equation, $\Sigma$ is the associated self energy, and ``Tr'' and ``Ln'' are extensions of ``tr'' and ``$\ln$'' to matrix-valued functions.  Equation~(\ref{pSigma}) is just the auxiliary field effective action~\cite{haymaker}, which yields the free fermion pressure in the absence of interactions; i.e., when $\Sigma \equiv 0$.  Owing to Eq.\,\,(\ref{Dkmu}), in this analysis we can neglect the gluon contribution to the pressure.

\begin{figure}[t]
\centerline{\includegraphics[scale=0.810] {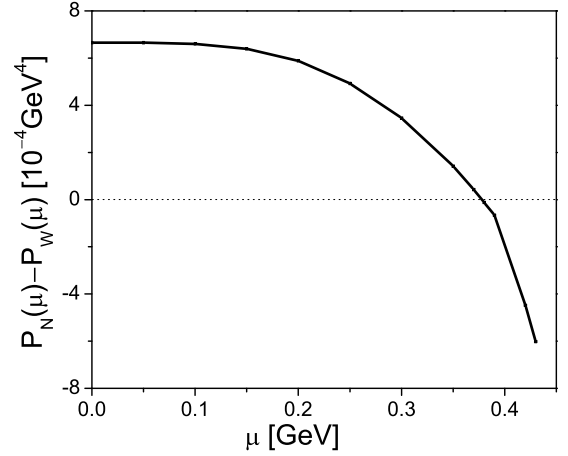}} 
\caption{Difference in pressure, Eq.\,(\protect\ref{bagconstant}), between the Nambu-Goldstone and Wigner solutions of the gap equation defined through Eqs.\,(\ref{KernelAnsatz}), (\ref{rainbowV}) and (\ref{IRGs}).  It is noteworthy that 
${\cal B}(0)=(0.16\,{\rm GeV})^4$ since a typical phenomenological value for bag constant is $\sim(0.15\,{\rm GeV})^4$. \label{fig:P(N-W)}}
\end{figure}

A system's ground state is that configuration for which the pressure is a global maximum or, equivalently, the effective-action is a global minimum.  In the steepest-descent approximation the pressure difference between the Nambu-Goldstone and Wigner phases is given by 
\begin{eqnarray}
\lefteqn{{\cal B}(\mu) =p_{\Sigma_{NG}}(\mu)-p_{\Sigma_W}(\mu)}\\
\nonumber
&=& 4N_c\int\frac{d^4p}{(2\pi^4)} \left\{ \ln\left[\frac{|\vec{p}|^2 A^2 + \tilde p_4^2 C^2 + B^2}{|\vec{p}|^2 A_W^2 + \tilde p_4^2 C_W^2}\right] \right. \\
&& \left. + |\vec{p}|^2 \left( \sigma_A^2 - \sigma_{A_W}^2 \right) 
+ \tilde p_4^2 \left( \sigma_C^2 - \sigma_{C_W}^2\right) \rule{0em}{3ex}\right\}, \label{bagconstant}
\end{eqnarray}
wherein the subscript $W$ distinguishes the solutions obtained in the Wigner phase from those obtained in the Nambu-Goldstone phase.  Our result for this chemical-potential-dependent \emph{bag constant} \cite{Cahill:1985mh} is depicted in Fig.\,\ref{fig:P(N-W)}.  It is evident that the Nambu-Goldstone phase is favoured on
\begin{equation}
\label{muchicritical}
\mu < \mu_c^{\chi\,{\rm rb}}, \; \mu_c^{\chi\,{\rm rb}} = 0.38\,{\rm GeV} < \mu_a^{\rm rb}.
\end{equation}
A quantitatively similar value of $\mu_c^{\chi\,{\rm rb}}$ is obtained, e.g., in the studies of Refs.\,\cite{Blaschke:1997bj,Bender:1997jf,Harada:1998zq}.  

The last of the observations in Eq.\,(\ref{muchicritical}) makes plain that in the model under consideration in this subsection an analytic continuation of the $\mu=0$ solutions of the gap equation is sufficient to completely determine the phase structure.  Moreover, in the chiral limit the transition is first-order because the vacuum quark condensate is zero in the chiral-limit Wigner phase and hence the condensate changes discontinuously at $\mu= \mu_c^{\chi\,{\rm rb}}$.  This, too, is in agreement with earlier DSE studies; e.g., Refs.\,\cite{Blaschke:1997bj,Bender:1997jf,Harada:1998zq,Bloch:1999vk}.

\begin{figure}[t]
\centerline{
\includegraphics[scale=0.810] {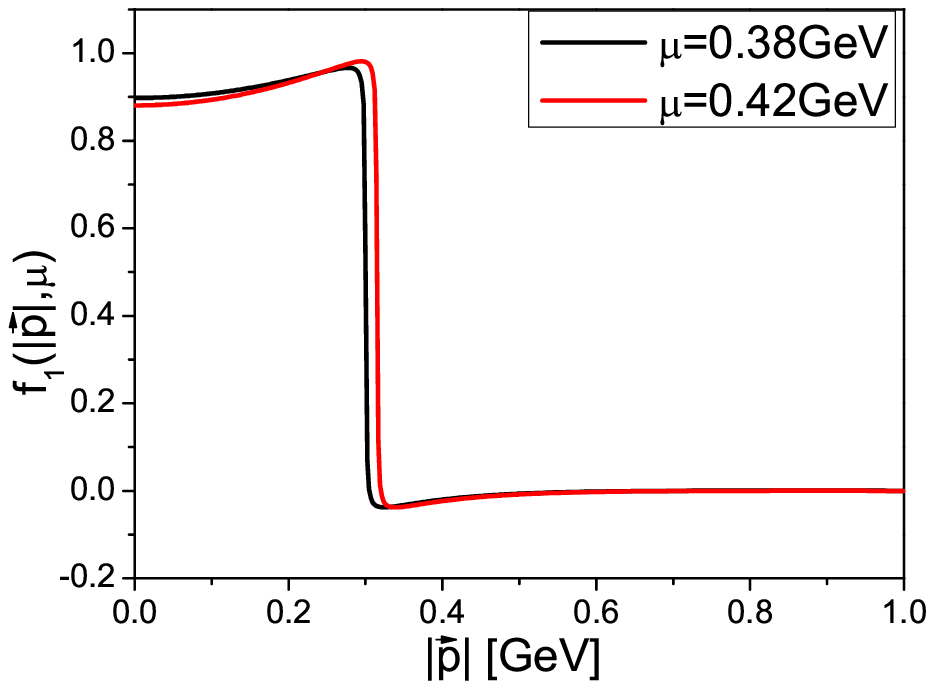}}  
\centerline{
\includegraphics[scale=0.810] {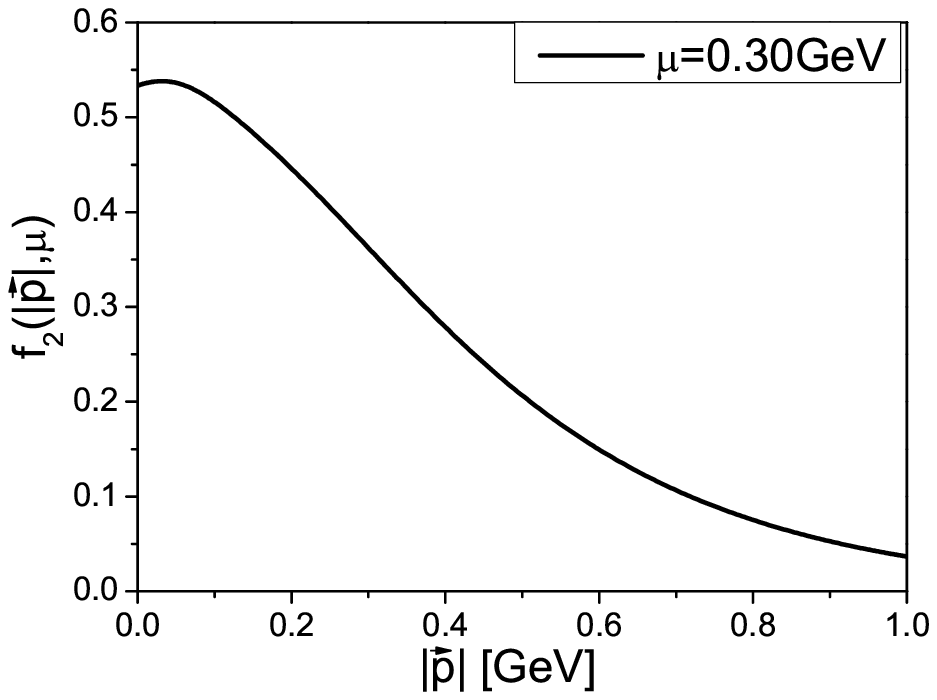}} 
\caption{\emph{Upper panel} -- Single-particle number-density distribution function, $f_1(|\vec{p}|,\mu)$ introduced in Eq.\,(\ref{nqmuf1}), computed at $\mu=0.38$, $0.42$ from the Wigner solution of the gap equation defined through Eqs.\,(\ref{KernelAnsatz}), (\ref{rainbowV}) and (\ref{IRGs}). This function is identically zero for $\mu<\mu_c^{\chi\,{\rm rb}}$.  \emph{Lower panel} -- Single-particle scalar-density distribution function, $f_2(|\vec{p}|,\mu)$ introduced in Eq.\,(\protect\ref{qbq0muf2}), computed at $\mu=0.3$ from the Nambu-Goldstone solution of the aforementioned gap equation.  This function is $\mu$-independent for $\mu<\mu_c^{\chi\,{\rm rb}}$ and identically zero otherwise. 
\label{fig:Dist-of-n}}
\end{figure}

It is notable that the derivative criteria explained in Sec.~2 of Ref.\,\cite{Roberts:2007ji} and exploited in Ref.\,\cite{BashirNew} can be employed\footnote{It is far easier to use this criterion than to compute the analogue of Eq.\,(\protect\ref{tauSchwinger}) obtained with $\sigma_{A_W}$.} to show that $\sigma_{A_W}(p)$ satisfies OS3 on $\mu\leq \mu_a$.   This is an indication that a first-order deconfinement transition is coincident with chiral symmetry restoration in the rainbow truncation model under consideration, which was also the case in the DSE studies of Refs.\,\cite{Blaschke:1997bj,Bender:1997jf,Bender:1996bm}.  

In the upper panel of Fig.\,\ref{fig:Dist-of-n} we plot the single-particle number-density distribution function introduced in Eq.\,(\ref{nqmuf1}).  It is identically zero for $\mu< \mu_c^{\chi\,{\rm rb}}<\mu_a$.  For $\mu>\mu_c^{\chi\,{\rm rb}}$ it is evaluated with the gap equation's Wigner solution and, since $\mu_c^{\chi\,{\rm rb}}>\mu_a^W$, the result is nonzero.  (The number density is discontinuous at $\mu_c^{\chi\,{\rm bc}}$.)  Evidently, on this domain the model supports an excess of quarks over antiquarks, a feature which is consistent with deconfinement because baryon bound states are not incorporated.  On the Wigner domain one has $A_W \neq C_W$.  Hence, as apparent in the figure, the Fermi momentum is 
\begin{equation}
p_F = \mu \, [C_W(0)/A_W(0)] \approx 0.7 \mu \,.  
\end{equation}
The pointwise behaviour of the curves in Fig.\,\ref{fig:Dist-of-n} is a nonperturbative property of the model's Wigner phase in rainbow truncation, not a numerical artefact.  Increased numerical precision does little more than sharpen the Fermi surface.  Comparison with Fig.\,\protect\ref{fig:Dist-of-n-BC} indicates features that are sensitive to the truncation.  The impact of a more realistic interaction \cite{Maris:1997tm,Maris:1999nt,Bloch:2002eq,Maris:2002mt,Eichmann:2008ae,Maris:2003vk} is currently being studied.

The lower panel of Fig.\,\ref{fig:Dist-of-n} shows the single-particle scalar-density distribution  function introduced in Eq.\,(\ref{qbq0muf2}).  It is independent of $\mu$ on the domain $\mu \leq\mu_c^{\chi\,{\rm rb}}$ but vanishes in the Wigner phase.  So long as  $\mu_c^{\chi}<\mu_a$ it is generally true that $f_2$ vanishes when $f_1$ is nonzero and vice versa.

The complete expression for the pressure will involve a contribution from colour-singlet baryon bound-states.  These are expressed in the gap equation via corrections to the rainbow truncation.  If a vertex \textit{Ansatz} is employed whose diagrammatic content is unknowable, as has often been done and we do subsequently, one cannot quantify the baryon contribution.  The genuine study of a baryon matter phase will only become possible when this situation is improved upon.  One anticipates it will yield a domain of chemical potential upon which the baryon number density and vacuum quark condensate are simultaneously nonzero.

\subsection{Dressed vertex}
\begin{figure}[t]
\centerline{
\includegraphics[scale=0.810] {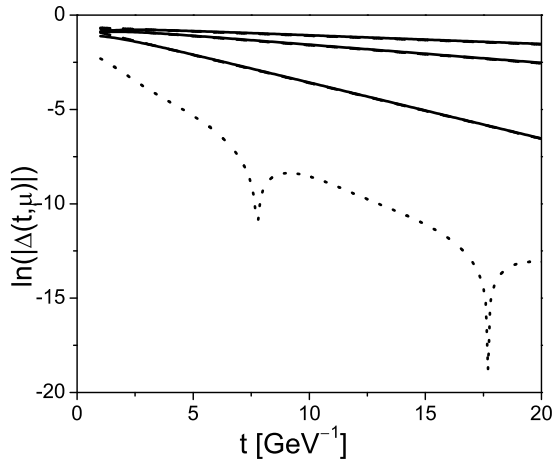}} 
\caption{Schwinger function in Eq.\,(\protect\ref{tauSchwinger}) computed from the gap equation defined via Eqs.\,(\ref{KernelAnsatz}), (\ref{bcvtx}) and (\ref{IRGs}).  \emph{Solid curves} -- top to bottom, $\mu = 0$, $0.2\,$GeV, $0.25\,$GeV.  \emph{Dotted curve} -- $\mu=0.31\,$GeV. \label{fig:SF-BC}}
\end{figure}

We now turn to a consideration of the dressed-vertex gap equation defined by Eqs.\,(\ref{KernelAnsatz}), (\ref{bcvtx}) and (\ref{IRGs}).  The in-vacuum solution is characterised by the quantities listed in Row~3 of Table~\ref{Table:Para1}.  In Fig.\,\ref{fig:SF-BC} we plot the Schwinger function defined in Eq.\,(\ref{tauSchwinger}), which is evidently in the class epitomised by Eq.\,(\ref{DeltaN}).  As usual the evolution with $\tau$ at small-$\mu$ is dominated by the pole for which $\Re(z)$ in Eq.\,(\ref{stinglSN}) is the smallest.  In this instance our fit yields $z_1 = M_1+i\Gamma_1$, with the values given in Table~\ref{Table:Para1}.  The quality of the fit is good and the absence of an imaginary part is noteworthy.  In this model, 
\begin{equation}
\mu_a^{\rm bc} = 0.30\,\,{\rm GeV}.
\end{equation}

\begin{figure}[t]
\centerline{\includegraphics[scale=0.810] {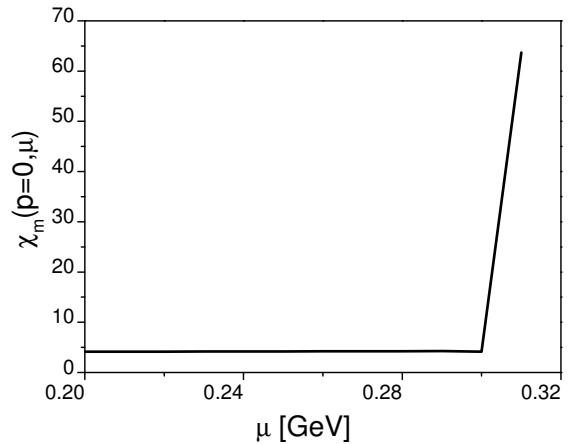}}
\centerline{\includegraphics[scale=0.810] {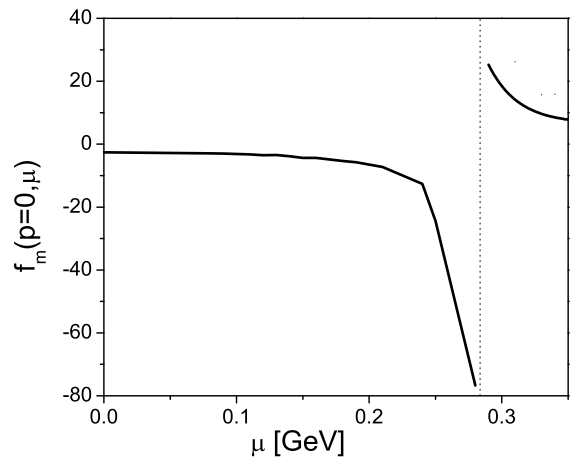}}  
\caption{Chiral susceptibilities defined in Eq.\,(\protect\ref{Xm}) and computed from the gap equation defined via Eqs.\,(\ref{KernelAnsatz}), (\ref{bcvtx}) and (\ref{IRGs}).  \emph{Upper panel} -- Nambu-Goldstone solution; \emph{lower panel} -- Wigner solution. \label{fig:chiralsus-BC}}
\end{figure}

Kindred to Fig.\,\ref{fig:SF1}, the pointwise evolution of the Schwinger function with $\tau$ is markedly different for $\mu\gtrsim\mu_a^{\rm bc}$.  It is consistent with the presence of at least one pair of additional singularities in Eq.\,(\ref{stinglSN}), which are characterised by $z_2 = M_2+i\Gamma_2$ with the fitted values listed in Table~\ref{Table:Para1}.  
For $\mu=0$ the presence of these other singularities does not necessarily entail that this dressed-quark Schwinger function violates the axiom of reflection positivity.  Hence, OS3 alone cannot be used to argue for quark confinement in this instance.

The gap equation associated with Eqs.\,(\ref{KernelAnsatz}), (\ref{bcvtx}) and (\ref{IRGs}) certainly admits solutions that exhibit chiral symmetry in the Nambu-Goldstone and Wigner modes.  However, owing to the fact that the diagrammatic content of Eq.\,(\ref{bcvtx}) is unknowable, it is impossible to write the truly corresponding dressed-quark contribution to the thermodynamic pressure.

How, then, is one to determine the critical chemical potential for chiral symmetry restoration, $\mu_c^{\chi\,{\rm bc}}$?  An estimate can be made by considering the chiral susceptibilities 
\begin{equation}
\label{Xm}
{\cal X}_m^P(\mu) 
= \left.\frac{\partial}{\partial m}  B^P(0;\mu)\right|_{\,m \simeq 0}, 
\end{equation}
where $m$ is the current-quark mass and also a source-term for $\langle \bar q q\rangle$, and $P= NG$ or $W$ indicates whether the scalar self-energy was calculated in the Nambu-Goldstone or Wigner phase \cite{Zhao:2006br}.\footnote{This procedure can work because the Wigner solution of the gap equation is accessible even for nonzero current-quark mass \protect\cite{Chang:2006bm}.}

In the neighbourhood of $\mu=0$ the Nambu-Goldstone phase is realised, a fact signalled by ${\cal X}_m^{NG}(\mu) >0$ and ${\cal X}_m^W(\mu) <0$.  The Nambu-Goldstone susceptibility is positive because the $B^{NG}(p;\mu)$ solution of the gap equation characterises the true vacuum and hence must be stable against a variation in the source-term for a chiral condensate.  ${\cal X}_m^{NG}$ will remain positive until a point $\mu=\mu_c^{NG}$, whereat the system becomes unstable against variations in the chiral-condensate source-term and ${\cal X}_m^{NG}(\mu)$ therefore exhibits a pole.  On the other hand, at small $\mu$ the Wigner susceptibility is negative because the Wigner phase is unstable against the formation of a quark condensate.  As $\mu$ is increased, ${\cal X}_m^W(\mu)$ will remain negative until a point $\mu=\mu_c^{W}$ whereat the Wigner vacuum becomes metastable and ${\cal X}_m^W(\mu)$ exhibits a pole.  Plainly, it must be that in any given theory
\begin{equation}
\mu_c^{W} < \mu_c^{\chi} < \mu_c^{NG}\,.
\end{equation}
Moreover, based on the results in extant model studies \cite{Zhao:2006br,Buballa:2003qv}, we anticipate that
\begin{equation}
\mu_c^{\chi} \approx \frac{\mu_c^{NG}+\mu_c^{W}}{2}\,.
\end{equation}
As an additional example, for the rainbow truncation described in Sect.\,\protect\ref{RainbowPT} this expression predicts $\mu_c^{\chi} = (0.54+0.24)/2=0.39\,$GeV, which compares well with Eq.\,(\protect\ref{muchicritical}).

In Fig.\,\ref{fig:chiralsus-BC} we plot the susceptibilities defined in Eq.\,(\ref{Xm}) and computed from the gap equation defined via Eqs.\,(\ref{KernelAnsatz}), (\ref{bcvtx}) and (\ref{IRGs}).  They exhibit the anticipated behaviour and enable us to infer
\begin{equation}
\label{muchibc}
\mu_c^{\chi\,{\rm bc}} \approx 0.3\,{\rm GeV} \approx \mu_a^{\rm bc}\,.
\end{equation}
Evidently, in the dressed-vertex model we cannot at present separate $\mu_c^{\chi\,{\rm bc}}$ from $\mu_a^{\rm bc}$.  Hereafter we will assume these points to be coincident.

\begin{figure}[t]
\centerline{
\includegraphics[scale=0.810] {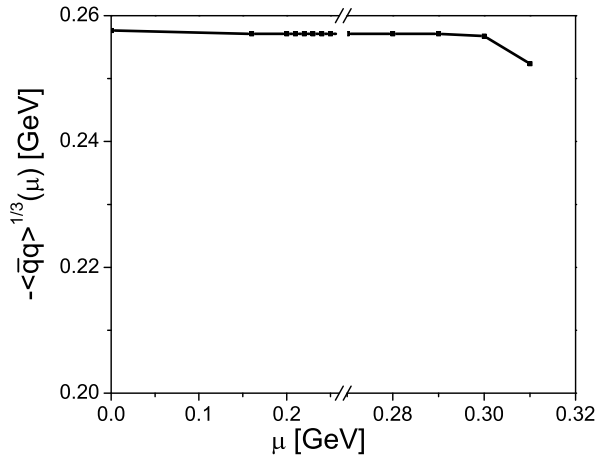}} 
\caption{Chemical potential dependence of the chiral-limit vacuum quark condensate computed from the gap equation defined via Eqs.\,(\ref{KernelAnsatz}), (\ref{bcvtx}) and (\ref{IRGs}).  Recall that $\mu_a^{\rm bc}=0.30\,$GeV. \label{fig:condensate-BC}}
\end{figure}

\begin{figure}[htb]
\centerline{
\includegraphics[scale=0.810] {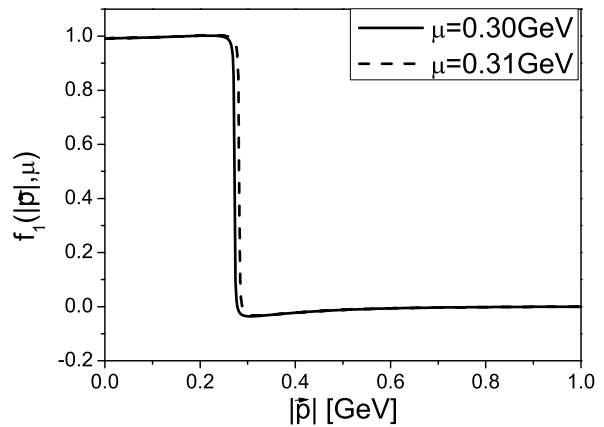}} 
\centerline{
\includegraphics[scale=0.810] {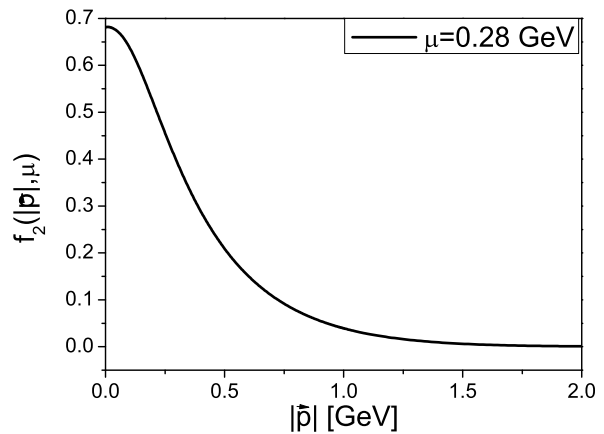}} 
\caption{\emph{Upper panel} -- Single-fermion number-density distribution function, $f_1(|\vec{p}|,\mu$, computed from the Wigner solution of the gap equation defined via Eqs.\,(\ref{KernelAnsatz}), (\ref{bcvtx}) and (\ref{IRGs}); \emph{lower panel} -- single-fermion scalar-density distribution function computed from this gap equation. \label{fig:Dist-of-n-BC}}
\end{figure}

The evolution with chemical potential of the chiral-limit dressed-vertex vacuum quark condensate is depicted in Fig.\,\ref{fig:condensate-BC}.  It is constant for $\mu < \mu_c^{\chi\,{\rm bc}}$ and vanishes for $\mu > \mu_c^{\chi\,{\rm bc}}$.  The chiral symmetry restoring transition is therefore first-order. 
Employing the derivative criteria explained in Sec.~2 of Ref.\,\cite{Roberts:2007ji} and employed in Ref.\,\cite{BashirNew}, we find in the dressed-vertex case, too, that $\sigma_{A_W}(p)$ satisfies OS3 on $\mu\leq \mu_a$.  

In the top panel of Fig.\,\ref{fig:Dist-of-n-BC} we plot the chiral-limit dressed-vertex fermion number-density distribution function.  It is only nonzero in the Wigner phase; i.e., for $\mu>\mu_c^{\chi\,{\rm bc}}$ in Eq.\,(\ref{muchibc}).  Therefore on this domain the pressure receives a contribution from an excess of quarks over antiquarks.  This is consistent with deconfinement but more cannot reliably be said because the diagrammatic content of the vertex \emph{Ansatz} is unknowable.

In the lower panel of Fig.\,\ref{fig:Dist-of-n-BC} we depict the chiral-limit fermion scalar-density distribution function.  It is independent of chemical potential for $\mu<\mu_c^{\chi\,{\rm bc}}$ and is only nonzero in the Nambu-Goldstone phase. 

\bigskip

\section{Summary and Conclusions}
\label{sec:summary}
The gap equation is the natural starting point for continuum studies of chiral symmetry restoration at nonzero chemical potential; viz., $\mu\neq 0$.  We have explored and illustrated its features using two models for the equation's kernel that are distinguished from each other by the form of the vertex \textit{Ansatz}.  It is important that qualitatively our results are largely insensitive to the \textit{Ansatz}.

In general the gap equation's kernel possesses a domain of analyticity in $\mu$.  This means that there exists a $\mu_a$ such that for all $\mu< \mu_a$ the gap equation's solution is simply obtained from the in-vacuum result through analytic continuation.  While examples exist in which $\mu_a = 0$, that is not necessarily the case and hence this observation can be practically useful.  For example, it guarantees that when calculated with the gap equation solution which expresses dynamical chiral symmetry breaking (DCSB), the single-quark number- and scalar-density distribution functions are $\mu$-independent on $\mu< \mu_a$.  This will also be true of other physical quantities.

A striking signature of dynamical chiral symmetry breaking is a nonzero value of the in-vacuum dressed-quark mass function, $M(p^2)$, which entails a nonzero chiral-limit vacuum quark condensate.  Both models possess this feature and realise chiral symmetry in the Nambu-Goldstone mode.  Naturally, a nonzero chemical potential destabilises the quark condensate and, with increasing chemical potential, both models exhibit a first-order chiral symmetry restoring transition at $\mu \approx M(0)$.  We described evidence which suggests the existence of a coincident first-order deconfinement transition.  

The models we studied are novel examples of the case $M(0)\approx \mu_c^\chi \leq \mu_a$.  However, they are alike flawed in the ultraviolet behaviour of the kernels.  This can be remedied by incorporating a term in the effective interaction which ensures that the one-loop renormalisation group behaviour of QCD is respected.  A kernel of this type has been widely employed in the study of hadron properties in-vacuum \cite{Maris:1997tm,Maris:1999nt,Bloch:2002eq,Maris:2002mt,Eichmann:2008ae,Maris:2003vk}.  An exploration of the $\mu\neq 0$ properties of such a model is underway.  

\begin{acknowledgments}
We are grateful for thoughtful comments from I.\,C.~Clo\"et, G.~Eichmann, B.~El-Bennich, P.~Jaikumar, A.~Krassnigg and R.\,D.~Young.
This work was supported by: 
the Department of Energy, Office of Nuclear Physics, contract no.\ DE-AC02-06CH11357;
the National Natural Science Foundation of China under Contract Nos.\ 10425521, 10575005, 10675007 and 10705002;
the Major State Basic Research Development Program under Contract No.\ G2007CB815000;
the Key Grant Project of the Chinese Ministry of Education under contact No.\ 305001;
and the Research Fund for the Doctoral Program of Higher Education of China under
grant No.\ 20040001010.  
\end{acknowledgments}

\end{document}